\documentclass[10pt,showpacs,amsmath,amssymb]{revtex4}

\usepackage{times}
\usepackage{graphicx}% Include figure files
\usepackage{bbold}
\usepackage{epsfig}
\usepackage{rotating}
\usepackage{graphpap}

\usepackage[]{graphicx}
\chardef\bslash=`\\ % p. 424, TeXbook

\begin{document}
\setlength{\unitlength}{1mm}

%%
%% issueinfo for header and copyright line
%%
%% \Dateposted{3 December 1900}
%%
%%
\pacs{05.45.-a, 87.18.Hf, 89.75.Fb\\[0.3cm]
{\it Dedicated to Hermann Haken on the occasion of his 80th birthday.}}

\title{Synergetic Analysis of the H{\"a}ussler-von der Malsburg Equations\\ for Manifolds of Arbitrary Geometry}

\author{M. G{\"u}{\ss}mann\footnote{E-mail: {\sf martin.guessmann@itp1.uni-stuttgart.de}}}
\affiliation{1.$\,$Institut f{\"u}r Theoretische Physik, Universit{\"a}t Stuttgart, Pfaffenwaldring 57, 70569 Stuttgart, Germany}
\author{A. Pelster\footnote{Corresponding author \quad E-mail: {\sf axel.pelster@uni-due.de}}}
\affiliation{Fachbereich Physik, Campus Duisburg, Universit{\"a}t Duisburg-Essen, Lotharstrasse 1, 47048 Duisburg, Germany}
\author{G. Wunner\footnote{E-mail: {\sf guenter.wunner@itp1.uni-stuttgart.de}}}
\affiliation{1.$\,\,$Institut f{\"u}r Theoretische Physik, Universit{\"a}t Stuttgart, Pfaffenwaldring 57, 70569 Stuttgart, Germany}

\begin{abstract}
We generalize a model of H{\"a}ussler and von der Malsburg which describes the 
self-organized generation of retinotopic 
projections between two one-dimensional discrete cell arrays on the basis of cooperative and competitive interactions of 
the individual synaptic contacts. 
Our generalized model is independent of the special geometry of the cell arrays and 
describes the temporal evolution of the connection weights between cells on different 
manifolds. By linearizing the equations 
of evolution around the stationary uniform state 
we determine the critical global growth rate for synapses onto the tectum 
where an instability arises. 
Within a nonlinear analysis we use then the methods of synergetics 
to adiabatically eliminate the stable modes near the instability. The resulting order parameter 
equations describe the emergence of retinotopic projections from initially undifferentiated 
mappings independent of dimension and geometry.
\end{abstract}
\maketitle
\section{Introduction}
\begin{figure}[t!]
\centerline{
\setlength{\unitlength}{1mm}
\includegraphics[width=11cm]{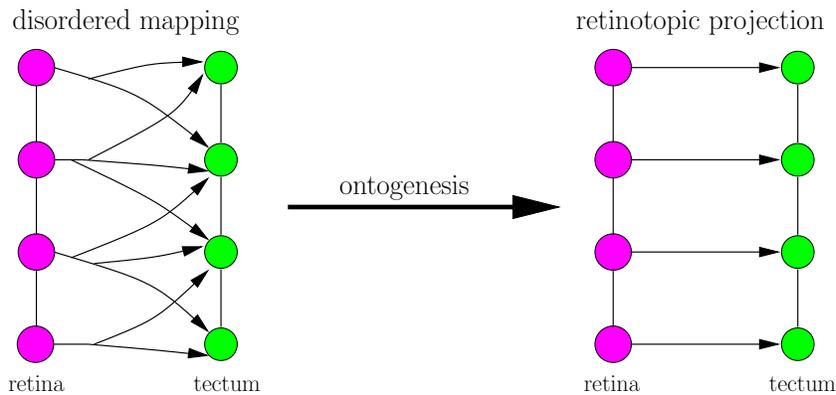}}
\caption{\label{FIG1}
In the course of ontogenesis originally disordered mappings between retina and tectum evolve 
into ordered projections.}
\end{figure}

An important part of the visual system of vertebrate animals are the neural connections 
between the eye and the brain.
At an initial stage of ontogenesis the ganglion cells of the retina have random synaptic 
contacts with the tectum,
a part of the brain which plays an important role in processing optical information. In the 
adult animal, however,
neighboring retinal cells project onto neighboring cells of the tectum (see Figure 
\ref{FIG1}). 
Further examples of these so-called {\it retinotopic} projections are established between the 
retina and
the corpus geniculatum laterale as well as the
visual cortex, respectively \cite{kandel}. This
conservation of neighborhood relations is also realized in many other neural connections 
between different cell sheets.
For instance, the formation of ordered projections between the mechanical receptors in the 
skin and the somatosensorial cortex
is called somatotopy. An even more abstract topological projection arises when 
the spatially resolved detection of similar frequencies in the ear
are projected onto neighboring cells of the auditorial cortex. 
A further notable neural map in the auditory system was discovered in the brain of the owl,
where neighboring cells of the {\it Nucleus mesencephalicus lateralis dorsalis} (MLD)
are excited by neighboring space areas, i.e.~every space point is represented by a small zone of the MLD \cite{knudsen}.
The variety of examples suggest 
that there must be some
underlying general mechanism for rearranging the initially disordered synaptic contacts into topological 
projections.\\

In the early 1940s, Sperry performed a series of pioneering experiments 
in the visual system of frogs and goldfish \cite{sperry1,sperry2}. Fish and amphibians can regenerate 
axonal tracts in their central nervous system, in contrast to mammals, birds 
and reptiles. Sperry crushed the optical nerve and found that retinal axons 
reestablished the previous retinotopically ordered pattern of connections in the 
tectum. Then in the early 
1960s Sperry presented his chemoaffinity hypothesis which proposed that the 
retinotectal map is set up on the basis of chemical markers carried by the cells \cite{sperry3}. 
However, experiments over several decades have shown that the formation of
retinotectal maps cannot be explained by this gradient matching alone \cite{goodhill}.\\ 

The group of von der Malsburg suggested that these ontogenetic processes result 
from self-organization.
The basic notion in their theory is the following: Once a fibre has already
grown from the retina to the tectum, the fibre moves along by strengthening its contacts in some parts 
of its ramification and by
weakening them in others. It is assumed that these modifications are governed by two 
contradictory 
rules \cite{Malsburg1,Malsburg2}:
on the one hand, synaptic contacts on neighboring tectal cells stemming from fibres of the same 
retinal region support each other to be
strengthened. On the other hand, the contacts starting from one retinal cell or ending at one
tectal cell compete with each other. In the case that 
retina and tectum are treated as one-dimensional discrete cell arrays, extensive computer 
simulations have shown that
a system based on these ideas of cooperativity and competition establishes, indeed, retinotopy 
as the final 
configuration \cite{Malsburg2}.
This finding was confirmed by a detailed analytical treatment of H{\"a}ussler and von der Malsburg 
\cite{Malsburg3} where the
self-organized formation of the synaptic connections between retina and tectum is described 
by an appropriate
system of ordinary differential equations. Applying the methods of synergetics 
\cite{Haken1,Haken2} for one-dimensional discrete cell arrays,
they succeeded in classifying the possible retinotopic projections and to discuss the criteria 
which determine their emergence. 
The more complicated case of continuously distributed cells on a spherical shell was partially 
discussed in 
Ref.~\cite{Malsburg4}.\\

It is the purpose of this paper to follow the outline of Ref.~\cite{Bochum} and generalize the original approach 
by elaborating a model for the self-organized formation of retinotopic projections
which is {\it independent} of the special geometry and dimension of the cell sheets. 
There are three essential reasons which motivate this more general approach. First, 
neurons usually do not establish 1-dimensional arrays but 2- or 3-dimensional networks. 
Hence the 1-dimensional model of H{\"a}ussler and von der Malsburg can only serve as a simplistic approximation of the real situation. 
Secondly, we want to include cell sheets of different extent, which is a more 
realistic assumption than neural sheets with the same number of cells. The third reason is that a general model is able to reveal what is generic, 
i.e.~what is independent of the special geometry of the problem. 
Thus, here we generalize the H{\"a}ussler equations to continuous manifolds of arbitrary geometry.
By doing so, we proceed in a phenomenological manner and relegate a microscopic derivation of the underlying equations to future research.\\ 

It should be emphasized that our main objective is not 
the biological modelling of retinotopy. 
Instead of that our considerations are devoted to the analysis of the dynamics of the nonlinear
H{\"a}ussler equations by using mathematical methods from nonlinear dynamics and synergetics.
For the more biological aspects of retinotopy and the vast progress in modelling
various retinotopically ordered projections during the last twenty years we refer
the reader to the reviews~\cite{goodhill,goodhill2,swindale}.\\ 

In Section \ref{SECMO} 
we present the general framework of our
model and introduce the equations of evolution for the connection weights between retina and 
tectum. We then perform
in Section \ref{SECLI} a linear stability analysis for the equations of evolution around
the stationary uniform state and discuss under which circumstances an 
instability arises. In Section \ref{NONLI} we apply the methods of synergetics,
and elaborate within a nonlinear analysis that the adiabatic 
elimination of the fast 
evolving degrees of
freedom leads to effective equations of evolution for the slow evolving order parameters. They 
approximately describe the dynamics near the instability
where an increase of the uniform growth rate of new synapses onto the tectum beyond a 
critical value converts an initially
disordered mapping into a retinotopic projection.
Finally, Section~\ref{summary} and \ref{outlook} provide a summary and an outlook. 
\section{General Model} \label{SECMO}
In this section we summarize the basic assumptions of our general model.
\subsection{Manifolds and Their Properties} \label{Manifolds}
We start with representing retina ($R$) and tectum ($T$) by general manifolds 
${\cal M}_T$ and ${\cal M}_R$, respectively. 
In the framework of an embedding of these manifolds in an Euclidean space of 
dimension $D$,  
the coordinates $x_R$, $x_T$ of the corresponding cells can be represented by
\begin{equation} \label{xTxR}
x_R=(x_R^1,x_R^2,\ldots,x_R^D)\,,\qquad x_R\in {\cal M}_R\,; \hspace*{1cm}
x_T=(x_T^1,x_T^2,\ldots,x_T^D)\,,\qquad x_T\in {\cal M}_T \, .
\end{equation}
In the following we need measures of distance, i.e.~metrics $g_{\mu\nu}^R$, $g_{\mu\nu}^T$ 
on the manifolds. The intrinsic coordinates of the $d$-dimensional manifolds 
${\cal M}_R$, ${\cal M}_T$ are denoted by $r^{\mu}$, $t^{\mu}$. Thus, the vectors (\ref{xTxR}) of the Euclidean embedding space can be parametrized according to
$x_R=x_R(r^{\mu})$, $x_T=x_T(t^{\mu})$.
With the covariant metric tensors 
\begin{equation} \label{metrtensor}
g_{\mu\nu}^R=\frac{\partial x_R}{\partial r^{\mu}}\,\frac{\partial x_R}{\partial r^{\nu}}\,,\quad g_{\mu\nu}^T=\frac{\partial x_T}{\partial t^{\mu}}\,\frac{\partial x_T}{\partial t^{\nu}}
\end{equation}
the line elements on the manifolds are given by
$(ds_R)^2=g_{\mu\nu}^R dr^{\mu}dr^{\nu}\,,\, (ds_T)^2=g_{\mu\nu}^T dt^{\mu}dt^{\nu}\,.$
The geodetic distances between two points of the manifolds read
\begin{equation} \label{geodabst}
s_{r r'}^R=\int\limits_{r'}^{r}\!\!\sqrt{g_{\mu\nu}^R\,dr^{\mu}dr^{\nu}}\,,\qquad 
s_{t t'}^T=\int\limits_{t'}^{t}\!\!\sqrt{g_{\mu\nu}^T\,dt^{\mu}dt^{\nu}}\,.
\end{equation}
We define a measure for the magnitudes of the manifolds by
\begin{equation}
\label{M1}
M_T=\int\! dt\,,\qquad M_R =\int\! dr\,,
\end{equation}
where we integrate over all elements of ${\cal M}_T$, ${\cal M}_R$. 
We characterize the neural connectivity within each manifold ${\cal M}_T$, ${\cal M}_R$ 
by cooperativity functions $c_T(t,t')$,
$c_R(r,r')$. In lack of any theory for the cooperativity functions
we regard them as time-independent, given properties of the manifolds
which are only limited by certain global plausible constraints. We assume 
that the cooperativity functions are positive
\begin{equation} 
\label{cpos}
c_T(t,t')\geq 0\,,\qquad c_R(r,r')\geq 0\,,
\end{equation}
that they are symmetric with respect to their arguments
\begin{equation} \label{csym}
c_T(t,t')=c_T(t',t),\qquad c_R(r,r')=c_R(r',r)\,,
\end{equation}
and that they fulfill the normalization conditions
\begin{equation} 
\label{NORM}
\int\! dt'\, c_T(t,t')=1,\qquad 
\int\! dr'\, c_R(r,r')=1\,.
\end{equation}
Furthermore, it is neurophysiologically reasonable to assume that the cooperativity 
functions $c_T(t,t')$,
$c_R(r, r')$ are larger when the distance between 
the points $t,t'$ and $r,r'$ is smaller. This condition of monotonically decreasing 
cooperativity functions can be written as
\begin{equation}
\label{DIST}
c_T(t,t')>c_T(t,t'')\,\mbox{ if }\,(s_{t t'}^T)^2<(s_{t t''}^T)^2\,,\quad
c_R(r,r')>c_R(r,r'')\,\mbox{ if }\,(s_{r r'}^R)^2<(s_{r r''}^R)^2\,.
\end{equation}

\subsection{Equations of Evolution}
\begin{figure}[t!]
 \centerline{a)\includegraphics[scale=0.52]{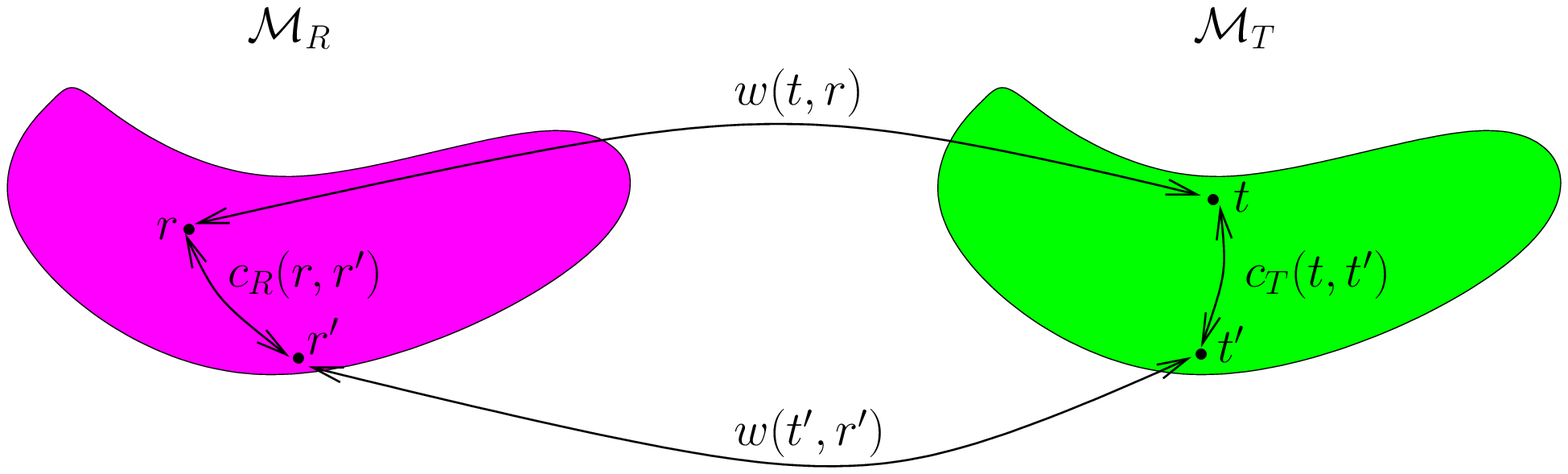}}\vspace{1cm}
 \centerline{b)\includegraphics[scale=0.52]{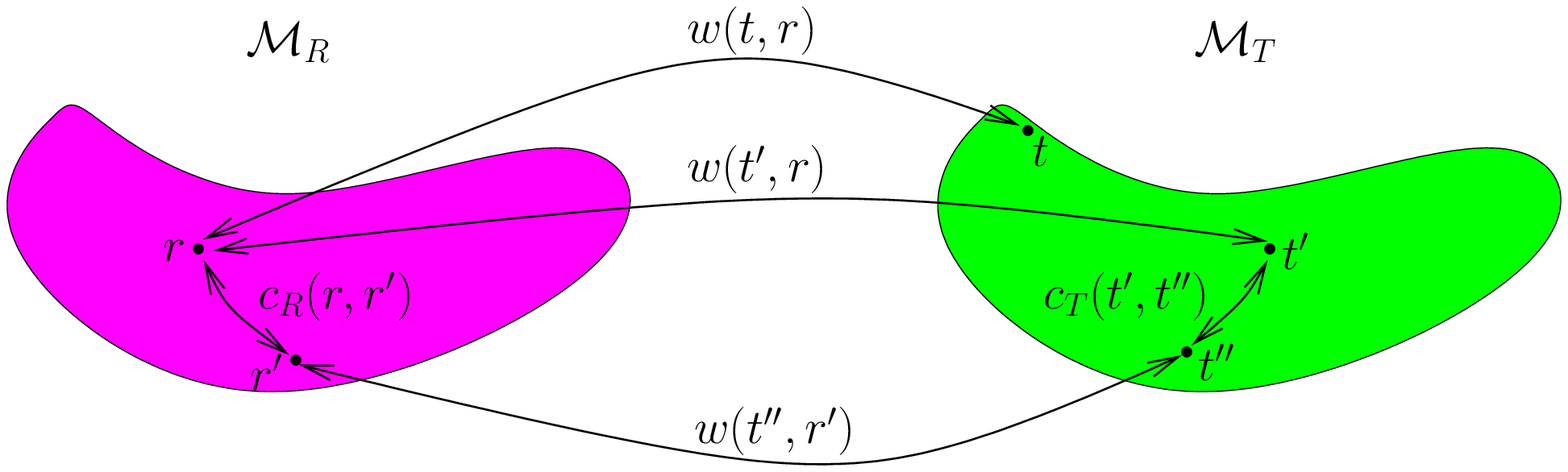}}\vspace{1cm}
 \centerline{c)\includegraphics[scale=0.52]{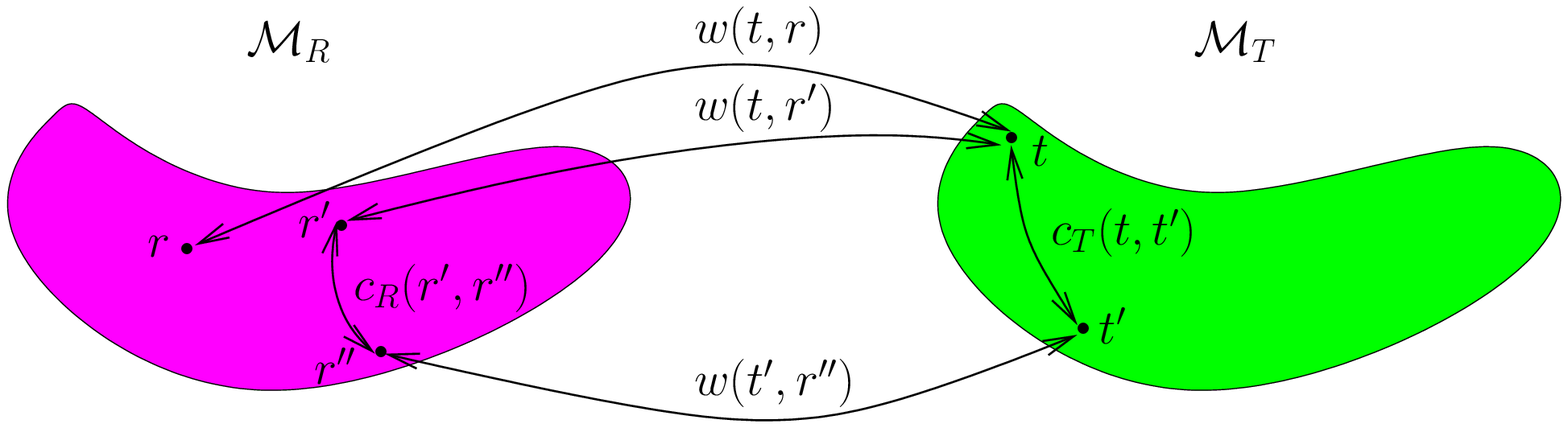}}
 \caption{\label{FIG2} \small Illustrations for the respective contributions to the generalized H{\"a}ussler equations (\ref{HAUSB}). Discussion see text.}
\end{figure}

The neural connections between retina and tectum are described by a connection weight 
$w(t,r)$ for
every ordered pair $(t,r)$ with $t\in {\cal M}_T,
\,r\in {\cal M}_R$. In this paper we 
are interested in
the temporal evolution of the connection weight $w(t,r)$ which is essentially 
determined by the given cooperativity
functions $c_T(t,t')$, $c_R(r,r')$ of the manifolds ${\cal M}_T$, 
${\cal M}_R$. To this end we generalize
a former ansatz of H{\"a}ussler and von der Malsburg \cite{Malsburg3} and assume that the 
evolution is governed by the following system of
ordinary differential equations \cite{thesis}:
\begin{eqnarray}
\dot w(t,r)&=&\alpha+w(t,r)\,\int\! dt' \,
\int\! dr' c_T(t,t')\,c_R(r,r')
\,w(t',r') \nonumber\\
&& -\frac{w(t,r)}{2 M_T}\hspace*{1mm}\int\! dt'
\left[ \alpha+w(t',r)\,\int\! dt''\,\int\! dr'
c_T(t',t'')\,c_R(r,r')\,w(t'',r')\right] \nonumber\\
&& -\frac{w(t,r)}{2M_R}\hspace*{1mm}\int\! dr'\left[ \alpha+w(t,r')
\,\int\! dt' \, \int\! dr'' c_T(t,t')\,
c_R(r',r'')\,w(t',r'')\right]  \,.
\label{HAUS}
\end{eqnarray}
Here $\alpha$ denotes the uniform growth-rate of new synapses onto the tectum, which will be 
the control parameter
of our system. These equations of evolution represent a balance
between  different cooperating and competing processes. To see this, we define the growth 
rate between the cells 
at $r$ and $t$
\begin{equation}
\label{GRO}
f(t,r,w)=\alpha+w(t,r) \int\! dt'\,
\int\! dr' c_T(t,t')\,
c_R(r,r')\,w(t',r') \, ,
\end{equation}
so that the generalized H{\"a}ussler equations (\ref{HAUS}) reduce to
\begin{equation} 
\label{HAUSB}
\dot w(t,r)=f(t,r,w)-\frac{w(t,r)}{2 M_T} 
\hspace*{2mm} \int\! dt' f(t',r,w)
-\frac{w(t,r)}{2 M_R}\hspace*{2mm} 
\int\! dr' f(t,r',w)\,.
\end{equation}
The cooperative contribution of the connection between $r'$ and $t'$ to the 
growth rate between $r$ and 
$t$ is given by the product $w(t,r) c_T(t,t') 
c_R(r,r') w(t',r')$
as shown in Figure \ref{FIG2}a. Therefore, this cooperative contribution is integrated 
with respect to $r'$, $t'$
and added to the uniform growth rate $\alpha$ to yield the total growth rate (\ref{GRO}) between 
$r$ and $t$. Apart from this cooperative term in the equations of evolution 
(\ref{HAUSB}), the remaining terms 
describe competitive processes. The second term accounts for the fact that growth rates 
between $r$ and $t'$
compete with the connections between $r$ and $t$ (see Figure \ref{FIG2}b). 
Correspondingly, the third term describes
the competition of the growth rates between $r'$ and $t$ with the connections 
between $r$ and $t$ 
(see Figure \ref{FIG2}c).
\subsection{Lower Limits for the Connection Strength}
Now we show that the evolution of the system due to the generalized H{\"a}ussler equations (\ref{HAUS}) 
leads to a lower bound for the
connection weight. To this end we assume the inequality
\begin{equation}
\label{INE}
0\leq w(t,r) \leq W
\end{equation}
to be fulfilled for some initial configuration. Then we conclude that the quantity
\begin{equation}
\label{CG}
C(t,r,w)=\int\! dt'
\int\! dr' c_T(t,t')\,c_R(r,r')
\,w(t',r')
\end{equation}
is positive as both the cooperativity functions $c_T(t,t')$, 
$c_R(r,r')$ and the
connection weight $w(t',r')$ are positive due to (\ref{cpos}) and (\ref{INE}).
On the other hand we read off from the normalization of the cooperativity functions 
(\ref{NORM}) that $C(t,r,w)$
cannot be larger than $W$: $0\leq C(t,r,w)\leq W$.
With this we can find a lower bound for $\dot w(t,r)$ as follows. The 
growth rate (\ref{GRO}) reads together 
with (\ref{CG}):
$f(t,r,w)=\alpha+w(t,r)\,C(t,r,w)$.
It can be minimized by setting $C(t,r,w)=0$, i.e.
\begin{equation}
\label{MIN}
f(t,r,w)_{\rm min}=\alpha\,,
\end{equation}
whereas its maximum value follows from $C(t,r,w)=W$:
\begin{equation}
\label{MAX}
f(t,r,w)_{\rm max}=\alpha+W^2 \,.
\end{equation}
To obtain a lower bound for $\dot w(t,r)$ in the H{\"a}ussler equations (\ref{HAUSB}),
we insert the minimum (\ref{MIN}) of the growth rate
for the cooperative first term and its maximum (\ref{MAX}) for the remaining competitive terms: 
\begin{eqnarray}
\dot w(t,r)_{\rm min} &=&\alpha-w(t,r)\left(\alpha+W^2 \right)\,.
\end{eqnarray}
Hence a small but positive $w(t,r)$ is prevented by a positive rate $\alpha$ 
from becoming zero. 
In this way we can conclude that the 
connection weight $w(t,r)$
is positive, when the inequality (\ref{INE}) is valid in an initial 
configuration. All further investigations will
concentrate on solutions of the H{\"a}ussler equations (\ref{HAUS}) with
$w(t,r)\geq 0$. Note that, in particular, the growth
rates (\ref{GRO}) for such configurations are positive.
\subsection{Complete Orthonormal System}\label{COS}
To perform both a linear and a nonlinear analysis of the underlying 
H{\"a}ussler equations (\ref{HAUS}) we need
a complete orthonormal system for both manifolds ${\cal M}_T$ and ${\cal M}_R$. 
With the help of the contravariant components $g^{\lambda \mu}_T$, $g^{\lambda \mu}_R$ of the metric introduced in Section~\ref{Manifolds} we 
define the respective
Laplace-Beltrami operators on the manifolds 
\begin{eqnarray}
\label{LBO}
\Delta_T=\frac{1}{\sqrt{g_T}} \, \partial_{\lambda} \left( g^{\lambda\mu}_T \sqrt{g_T} \,\partial_{\mu}\right) \, , \hspace*{1cm}
\Delta_R=\frac{1}{\sqrt{g_R}}\, \partial_{\lambda}\left(g^{\lambda\mu}_R \sqrt{g_R}\, \partial_{\mu}\right) \, ,
\end{eqnarray}
where $g_T, g_R$ represent the determinants of the covariant components 
$g_{\lambda \mu}^T$, $g_{\lambda \mu}^R$ of the metric.
The Laplace-Beltrami operators 
allow to introduce a complete orthonormal system by their eigenfunctions 
$\psi_{\lambda_T} (t)$,
$\psi_{\lambda_R} (r)$ according to
\begin{eqnarray}
\label{LBO2}
\Delta_T \, \psi_{\lambda_T} (t)=\chi^T_{\lambda_T} \,\psi_{\lambda_T} (t) \, , \hspace*{1cm}
\Delta_R \, \psi_{\lambda_R} (r)=\chi^R_{\lambda_R} \,\psi_{\lambda_R} (r)\,.
\end{eqnarray}
Here $\lambda_T$, $\lambda_R$ denote discrete or continuous 
numbers which parameterize the eigenvalues $\chi^T_{\lambda_T}$, 
$\chi^R_{\lambda_R}$ of the Laplace-Beltrami operators which could be degenerate.
By construction, they fulfill the orthonormality relations
\begin{eqnarray}
\label{ORTHO}
\int\! dt\,\psi_{\lambda_T} (t) \psi_{\lambda_T'}^{*}(t)=
\delta_{\lambda_T \lambda_T'} \,, \hspace*{1cm}
\int\! dr\,\psi_{\lambda_R} (r) \psi_{\lambda_R'}^{*}(r)=
\delta_{\lambda_R \lambda_R'} \,,
\end{eqnarray}
and the completeness relations
\begin{eqnarray}
\label{COMP}
\sum_{\hspace*{2mm}_{\lambda_T}} 
\psi_{\lambda_T} (t)
\psi_{\lambda_T}^{*}(t')
=\delta(t-t')\,,\hspace*{1cm}
\sum_{\hspace*{2mm}{\lambda_R}}
\psi_{\lambda_R} (r) \psi_{\lambda_R}^{*}(r')
=\delta(r-r')\,.
\end{eqnarray}
Note that the explicit form (\ref{LBO}) of the Laplace-Beltrami operators 
enforces the eigenvalues 
$\chi^T_{\lambda_T=0}=0 \, , \chi^R_{\lambda_R=0}=0$ 
with the constant eigenfunctions
\begin{eqnarray}
\label{COO}
\psi_{\lambda_T=0} (t) = \frac{1}{\sqrt{M_T}} \, , \hspace*{1cm}
\psi_{\lambda_R=0} (r) = \frac{1}{\sqrt{M_R}}
\end{eqnarray}
because of (\ref{M1}) and the orthonormality relations (\ref{ORTHO}). 
The cooperativity functions can be expanded in terms of the eigenfunctions according to
\begin{equation}
c_T(t,t')=\sum_{\hspace*{2mm}{\lambda_T}}\sum_{\hspace*{2mm}{\lambda_T'}}
F_{\lambda_T \lambda_T '} \psi_{\lambda_T} (t) \psi_{\lambda_T'}^{*}(t')\,,\quad
c_R(r,r')=\sum_{\hspace*{2mm}{\lambda_R}}\sum_{\hspace*{2mm}{\lambda_R'}}
F_{\lambda_R \lambda_R'} \psi_{\lambda_R} (r) \psi_{\lambda_R'}^{*}(r')\,.
\end{equation}
In the following we assume for the sake of simplicity that the  
corresponding expansion coefficients are diagonal
$F_{\lambda_T \lambda_T'} = f_{\lambda_T} \delta_{\lambda_T \lambda_T'} \,,\,
F_{\lambda_R \lambda_R'} =f_{\lambda_R} \delta_{\lambda_R \lambda_R'} \,,$
so we have
\begin{eqnarray}
\label{CEX}
c_T(t,t')=\sum_{\hspace*{2mm}{\lambda_T}}
f_{\lambda_T} \psi_{\lambda_T} (t)
\psi_{\lambda_T}^{*}(t')\,,\hspace*{1cm}
c_R(r,r')=\sum_{\hspace*{2mm}{\lambda_R}}
f_{\lambda_R} \psi_{\lambda_R} (r)
\psi_{\lambda_R}^{*}(r')\,.
\end{eqnarray}
Thus, $\psi_{\lambda_T}(t)$, $\psi_{\lambda_R}(r)$ are not only eigenfunctions 
of the Laplace-Beltrami
operators as in (\ref{LBO2}) but also eigenfunctions of the cooperativity 
functions according to
\begin{eqnarray}
\int\! dt'\, c_T(t,t') \, \psi_{\lambda_T} (t') 
= f_{\lambda_T} \,\psi_{\lambda_T} (t) \, , \hspace*{1cm}
\int\! dr'\, c_R(r,r') \, \psi_{\lambda_R} (r')  
= f_{\lambda_R} \,\psi_{\lambda_R} (r) \, .
\end{eqnarray}
Note that the normalization of the cooperativity functions (\ref{NORM})  
and the orthonormalization relations (\ref{ORTHO}) lead to the constraints
$f_{\lambda_T=0}=f_{\lambda_R=0}= 1 \, .$
%
%Furthermore, we mention that the orthonormality relations (\ref{ORTHO}) allow 
%inverting the expansions (\ref{CEX}),
%so that the expansion coefficients $f_{\lambda_T}$, $f_{\lambda_R}$ can be 
%determined from the cooperativity functions 
%$c_T(t,t')$, $c_R(r,r')$
%
%\begin{equation}
%f_{\lambda_T}=\int\! dt\,\int\! dt'\,
%c_T(t,t')\,\psi_{\lambda_T}^{*}(t)
%\, \psi_{\lambda_T} (t') \, , \quad
%f_{\lambda_R} =\int\! dr\,\int\! dr'
%c_R(r,r')\,\psi_{\lambda_R}^{*}(r)
%\,\psi_{\lambda_R} (r')\,.
%\end{equation}
%
\section{Linear Stability Analysis}\label{SECLI}
\begin{figure}[t]
\setlength{\unitlength}{1mm}
\centerline{\includegraphics[width=12cm]{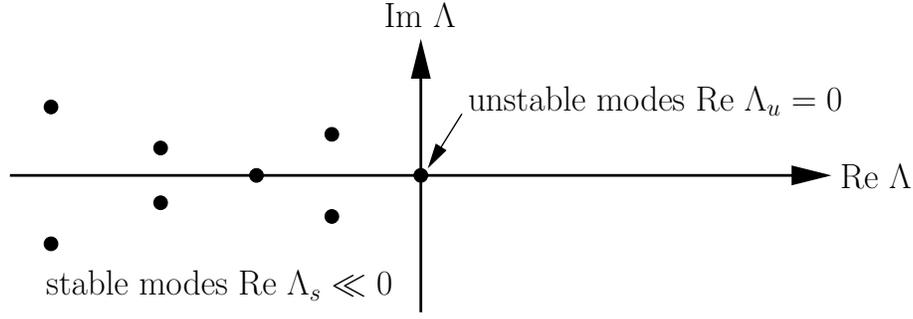}}
\caption{\label{FIG3}
Schematic representation of the eigenvalues (\ref{VAL}) 
at the instability. The unstable part consists of those eigenvalues
which nearly vanish whereas the stable part lies in a region separated by 
a finite distance from the stable part.}
\end{figure}

Now we employ the methods of synergetics \cite{Haken1,Haken2} and investigate 
the underlying equations of evolution (\ref{HAUS}) in the
vicinity of the stationary uniform solution. Inserting the ansatz 
$w(t,r)=w_0$ into the H{\"a}ussler equations 
(\ref{HAUS}), we take into account (\ref{M1})  
as well as the normalization
of the cooperativity functions (\ref{NORM}). By doing so, we deduce 
$w_0 = 1$. Let us introduce the deviation from this stationary uniform solution 
$v(t,r)=w(t,r)-1\,,$
and rewrite the H{\"a}ussler equations (\ref{HAUS}). Defining the linear operators
\begin{eqnarray}
\label{C} 
\hat{C} (t,r,x )&=&\int\! dt'  
\int\! dr'\,
c_T(t,t') \, c_R(r,r')\, x(t',r') \, , \\
\label{B}
\hat{B}(t,r,x )&=&
\frac{1}{2 M_T}
\int\! dt'\, x(t',r)+
\frac{1}{2 M_R}\int\! dr'\, x(t,r') \, ,
\end{eqnarray}
the resulting equations of evolution assume the form
\begin{equation} 
\label{gen43}
\dot v(t,r)=\hat{L}(t,r,v)
+ \hat{Q}(t,r,v )+\hat{K}(t,r,v ) \, .
\end{equation}
Here the linear, quadratic, and cubic terms, respectively, are given by
\begin{eqnarray}
\label{L17}
\hat{L}(t,r,v)&=&-\alpha v+\hat{C}(t,r,v)
-\hat{B}(t,r,v)-\hat{B}(t,r,\hat{C}(t,r,v)) \, , \label{L}\\
\hat{Q}(t,r,v)&=&v\, \left(\hat{C}(t,r,v)-\hat{B}(t,r,v )
-\hat{B}(t,r,\hat{C}(t,r,v)) \right) 
-\hat{B} \left(t,r,v\, \hat{C}(t,r,v)\right) \, ,\label{Q}\\
\hat{K}(t,r,v)&=&-v \, \hat{B}\left(t,r,v \, \hat{C}(t,r,v)\right)\label{K}\,.
\end{eqnarray}
To analyze the stability of the stationary uniform solution we neglect for the time being
the nonlinear terms in (\ref{gen43}) and investigate the linear problem
\begin{equation}
\label{LS} 
\dot v(t,r)=\hat{L}(t,r,v ) \, .
\end{equation}
Solutions of (\ref{LS}) depend exponentially on the time $\tau$,
$v(t,r) = v_{\lambda_T \lambda_R} (t,r) 
\exp \left( \Lambda_{\lambda_T \lambda_R} \, \tau \right)$
with $v_{\lambda_T \lambda_R}$ and $\Lambda_{\lambda_T \lambda_R}$ denoting the eigenfunctions and eigenvalues
of the linear  operator $\hat{L}$:
\begin{eqnarray}
\label{LEI}
\hat{L} \left(t,r,v_{\lambda_T \lambda_R}\right)=\Lambda_{\lambda_T \lambda_R}
\,v_{\lambda_T \lambda_R} (t,r)\, .
\end{eqnarray}
Now we use the complete
and orthonormal system on the manifolds ${\cal M}_T$, ${\cal M}_R$, 
which have been defined in Section \ref{COS}, and show that the
eigenfunctions of $\hat{L}$ are products of the form
\begin{equation} 
\label{EV}
v_{\lambda_T \lambda_R} (t,r)=\psi_{\lambda_T} (t)\,\psi_{\lambda_R} (r) \, .
\end{equation}
Indeed, when the operator (\ref{C}) acts on (\ref{EV}), 
the expansion of the cooperativity functions (\ref{CEX})
leads, together with the orthonormality relations (\ref{ORTHO}), to
\begin{eqnarray}
\hat{C}(t,r,v_{\lambda_T \lambda_R} )=
f_{\lambda_T}\,f_{\lambda_R}\,v_{\lambda_T \lambda_R} (t,r)
\,.
\label{EIG1}
\end{eqnarray}
Thus, the operator $\hat{C}$ has the eigenfunctions 
$v_{\lambda_T \lambda_R} (t,r)$ with
the eigenvalues $f_{\lambda_T} f_{\lambda_R}$. In a 
similar way we obtain for the operator (\ref{B}):
\begin{equation}
\hat{B}(t,r,v_{\lambda_T \lambda_R}) =\left\{\begin{array}{ccc}
v_{\lambda_T \lambda_R}&&\lambda_T=\lambda_R=0 \, ,\\*[2mm]
v_{\lambda_T \lambda_R}/2&\hspace*{0.5cm}&\lambda_T=0\,,\lambda_R\not=0;\lambda_R=0\,,\lambda_T\not=0\, ,\\*[2mm]
0&&\mbox{otherwise}\,.
\end{array} \right.
\label{EIG2}
\end{equation}
Combining the eigenvalue problems (\ref{EIG1}), (\ref{EIG2}) for $\hat{C}$ and $\hat{B}$, we find
\begin{eqnarray}
\hat{B}\left(t,r,\hat{C}(v_{\lambda_T \lambda_R})\right)=
\left\{\begin{array}{ccc}
f_{\lambda_T} f_{\lambda_R} v_{\lambda_T \lambda_R} &&\lambda_T=\lambda_R=0\vspace{0.15cm}\, ,\\
f_{\lambda_T} f_{\lambda_R} 
v_{\lambda_T \lambda_R}/2&\hspace*{0.5cm}&\lambda_T=0\,,\lambda_R\not=0;\lambda_R=0\,,\lambda_T\not=0\vspace{0.15cm}\, ,\\
0&&\mbox{otherwise}\,.
\end{array} \right.
\label{EIG3}
\end{eqnarray}
Thus, we conclude from (\ref{EIG1})--(\ref{EIG3}) 
that the linear operator $\hat{L}$ fulfills the eigenvalue
problem (\ref{LEI}) with the eigenfunctions (\ref{EV}) and the eigenvalues
\begin{equation}
\label{VAL}
\Lambda_{\lambda_T \lambda_R} =\left\{\begin{array}{ccc}
-\alpha-1&&\lambda_T=\lambda_R=0\vspace{0.15cm}\, ,\\
-\alpha + (f_{\lambda_T} f_{\lambda_R} -1)/2&\hspace*{0.5cm}
&\lambda_T=0\,,\lambda_R\not=0;\lambda_R=0\,,\lambda_T\not=0
\vspace{0.15cm}\, ,\\
-\alpha+f_{\lambda_T} f_{\lambda_R} &&\mbox{otherwise}\,.
\end{array} \right.
\end{equation}
By changing the uniform growth rate $\alpha$ in a suitable way, 
the real parts of some eigenvalues (\ref{VAL}) become positive
and the system can be driven to the neighborhood of an instability. 
Which eigenvalues (\ref{VAL}) become unstable
in general depends on the respective values
of the given expansion coefficients $f_{\lambda_T}$, $f_{\lambda_R}$. 
The situation simplifies, however, if we follow Ref. \cite{Malsburg3} 
and assume that the absolute values of the expansion coefficients 
$f_{\lambda_T}$, $f_{\lambda_R}$ are equal or smaller than the 
normalization value $f_0=1$:
$|f_{\lambda_T}|\leq 1 \, ,|f_{\lambda_R}|\leq 1$.
Then the eigenvalue in (\ref{VAL}) with the largest real part
is given by some parameters $\lambda_T^u, \lambda_R^u$ with
$\Lambda_{\rm max} = \Lambda_{\lambda_T^u \lambda_R^u} = - \alpha + f_{\lambda_T^u} f_{\lambda_R^u}$.
Thus, the linear stability analysis reveals that the instability arises 
at the critical uniform growth rate
\begin{eqnarray}
\label{INS}
\alpha_c =  \mbox{Re}\, (f_{\lambda_T^u} f_{\lambda_R^u}) 
\end{eqnarray}
and that its neighborhood is characterized by
\begin{eqnarray}
\mbox{Re}\, (\Lambda_{\lambda_T^u \lambda_R^u}) \approx 0\,;\hspace*{1cm}
\mbox{Re}\, (\Lambda_{\lambda_T \lambda_R}) \ll 0\,,\quad (\lambda_T;\lambda_R) \not = (\lambda_T^u;\lambda_R^u)\,.
\end{eqnarray}
Consequently, the absolute values of 
the eigenvalues of the unstable modes 
$(\lambda_T^u;\lambda_R^u)$ are much smaller than those of the stable modes 
$(\lambda_T;\lambda_R)\not=(\lambda_T^u;\lambda_R^u)$:
\begin{equation} 
\label{rea}
|\mbox{Re}\, (\Lambda_{\lambda_T^u \lambda_R^u})| 
\ll |\mbox{Re}\, (\Lambda_{\lambda_T \lambda_R})|\,,\qquad (\lambda_T;\lambda_R)\not=(\lambda_T^u;\lambda_R^u)\,.
\end{equation}
The resulting spectrum is schematically illustrated
in Figure~\ref{FIG3}. 
\section{Nonlinear Analysis} \label{NONLI}
In this section we perform a detailed nonlinear 
analysis of the H{\"a}ussler equations (\ref{HAUS}). Using the methods of
synergetics \cite{Haken1,Haken2} we derive our main result in form 
of the order parameter equations which describe the
emergence of retinotopic projections from initially undifferentiated mappings.
\subsection{Unstable and Stable Modes}
We return to the nonlinear equations of evolution (\ref{gen43}) 
for the deviation from the stationary uniform solution $v(t,r)$.
As the eigenfunctions $\psi_{\lambda_T}  (t)$, $\psi_{\lambda_R} (r)$ of the 
Laplace-Beltrami operators $\Delta_T$, $\Delta_R$ represent a complete orthonormal 
system on the manifolds ${\cal M}_T$, ${\cal M}_R$,
we can expand the deviation from the stationary solution according to
\begin{eqnarray}
\label{VEXP}
v(t,r) = \hspace*{2mm} 
V_{\lambda_T \lambda_R} \psi_{\lambda_T} (t) \psi_{\lambda_R} (r)\,.
\end{eqnarray}
Here we have introduced Einstein's sum convention, i.e.~repeated indices are implicitly summed over.
The sum convention is adopted throughout.
Motivated by the linear stability analysis of the preceding section, 
we decompose the expansion (\ref{VEXP}) near the instability
which is characterized by (\ref{INS}):
\begin{equation}
\label{US}
v(t,r) = U(t,r)+S(t,r)\,.
\end{equation}
We can expand the unstable modes in the form
\begin{equation}
\label{UEXP}
U(t,r)=
U_{\lambda_T^u \lambda_R^u} \psi_{\lambda_T^u} (t) \psi_{\lambda_R^u} (r) \, ,
\end{equation}
where the expansion amplitudes $U_{\lambda_T^u \lambda_R^u}$ 
will later represent the order parameters indicating the 
emergence of an instability. Correspondingly, 
\begin{equation}
\label{SEXP}
S(t,r)=
S_{\lambda_T \lambda_R}\, \psi_{\lambda_T} (t) \psi_{\lambda_R} (r)
\end{equation}
denotes the contribution of the stable modes. Note that the summation in (\ref{SEXP}) 
is performed over all parameters $(\lambda_T ;\lambda_R )$ 
except for $(\lambda_T^u ;\lambda_R^u )$,
i.e.~from now on the parameters $(\lambda_T;\lambda_R)$ stand for the stable modes alone. 
In the following we aim at deriving separate equations of 
evolution for the amplitudes $U_{\lambda_T^u \lambda_R^u}$, 
$S_{\lambda_T \lambda_R}$.
To this end we define the operators
\begin{eqnarray}
\hat P_{\lambda_T^u \lambda_R^u} (x)
&:=&\int\! dt
\int\! dr \,\psi_{\lambda_T^u}^{*}(t) \psi_{\lambda_R^u}^{*}(r) \,x (t,r) \, ,
\label{proj1}\\
\hat P_{\lambda_T \lambda_R} (x)  &:=&\int\! dt \int\! dr
\, \psi_{\lambda_T}^{*}(t) \psi_{\lambda_R}^{*}(r)\,x (t,r)\, ,
\quad  (\lambda_T;\lambda_R)\not=(\lambda_T^u;\lambda_R^u) \,,
\label{proj2}
\end{eqnarray}
which project, out of $v(t,r)$, the amplitudes  
of the unstable and stable modes, respectively:
$U_{\lambda_T^u \lambda_R^u}=\hat P_{\lambda_T^u \lambda_R^u} (v) \, ,$
$S_{\lambda_T \lambda_R}=\hat P_{\lambda_T \lambda_R} (v) \, .$
These equations follow from
(\ref{US})--(\ref{proj2}) by taking into account the orthonormality relations (\ref{ORTHO}).
With these projectors the nonlinear equations of evolution (\ref{gen43}) decompose into
\begin{eqnarray}
\dot U_{\lambda_T^u \lambda_R^u} &=& \Lambda_{\lambda_T^u \lambda_R^u} \, 
U_{\lambda_T^u \lambda_R^u}
+ \hat P_{\lambda_T^u \lambda_R^u} \left( \hat{Q}(t,r, U+S)\right)
+\hat P_{\lambda_T^u \lambda_R^u} \left( \hat{K}(t,r,U+S) \right) \, ,
\label{U}\\
\dot S_{\lambda_T \lambda_R} &=&\Lambda_{\lambda_T \lambda_R} S_{\lambda_T \lambda_R}
+\hat P_{\lambda_T \lambda_R} \left(
\hat{Q}(t,r,U+S) \right)+ 
\hat P_{\lambda_T \lambda_R} \left( \hat{K}(t,r,U+S)\right)\,. 
\label{S}
\end{eqnarray}
Note that we used the eigenvalue problem (\ref{LEI}) 
for the linear operator $\hat{L}$ and its eigenfunctions (\ref{EV})
to derive the first term on the right-hand side in (\ref{U}) and (\ref{S}),
where Einstein's sum convention is not applied.\\

In general, it appears impossible to determine a solution for
the coupled amplitude equations (\ref{U}), (\ref{S}). 
Near the instability which is characterized by (\ref{INS}),
however, the methods of synergetics \cite{Haken1,Haken2} allow 
elaborating an approximate solution which is based on the
inequality (\ref{rea}). To this end we interpret (\ref{rea}) in terms of a 
{\it time-scale hierarchy}, i.e.~the stable modes evolve on a faster 
time-scale than the unstable modes:
\begin{equation}
\tau_u=\frac{1}{|\mbox{Re}\, (\Lambda_{\lambda_T^u \lambda_R^u})|} \gg 
\tau_s=\frac{1}{|\mbox{Re}\, (\Lambda_{\lambda_T \lambda_R})|}\,.
\end{equation}
Due to this time-scale hierarchy the stable modes $S_{\lambda_T \lambda_R}$
quasi-instan\-taneously take values which are prescribed by the unstable 
modes $U_{\lambda_T^u \lambda_R^u}$.  
This is the content of the well-known {\it slaving principle} 
of synergetics: the stable modes are enslaved by the unstable modes. 
In our context it states mathematically that the dynamics of the stable modes  
$S_{\lambda_T \lambda_R}$ is determined by the center manifold $H$ according to
\begin{equation}
\label{H}
S_{\lambda_T \lambda_R} =H_{\lambda_T \lambda_R} \left(  U_{\lambda_T^u \lambda_R^u} \right)\,.
\end{equation}
Inserting (\ref{H}) in (\ref{S}) leads to an implicit equation for 
the center manifold $H$ which we approximately solve in the
vicinity of the instability below. By doing so, we adiabatically 
eliminate the stable modes from the relevant dynamics. Then we
use the center manifold $H$ in the equations of evolution (\ref{U}),  
i.e.~we reduce the original high-dimensional system to a low-dimensional one for the order parameters 
$U_{\lambda_T^u \lambda_R^u}$. The resulting
order parameter equations describe the dynamics near the instability 
where an increase of the uniform growth rate $\alpha$
beyond its critical value (\ref{INS}) converts disordered mappings into retinotopic projections.
\subsection{Integrals}
It turns out that the derivation of the order parameter equations 
contains integrals over products of eigenfunctions which have the form
\begin{equation} 
\label{abkurz}
I^{\lambda}_{\lambda^{(1)} \lambda^{(2)} \ldots  \lambda^{(n)}}
=\hspace*{1mm}\int\! dx\,
\psi_{\lambda}^{*}(x)\,\psi_{\lambda^{(1)}}(x)\,\psi_{\lambda^{(2)}}(x)\,\cdots\,
\psi_{\lambda^{(n)}}(x)\,,
\end{equation}
where $\lambda$, $x$ stand for the respective quantities $\lambda_T$, $t$  
and $\lambda_R$, $r$  of the manifolds ${\cal M}_T$ and ${\cal M}_R$. 
Examples for such integrals are:
\begin{equation}
I^{\lambda} =\int\! dx\,\psi_{\lambda}^{*}  (x) \, , 
\,\,\,I^{\lambda}_{\lambda'} =\int\! dx\,
\psi_{\lambda}^{*}(x)\,\psi_{\lambda'}(x)\,,
\,\,\,I^{\lambda}_{\lambda' \lambda''} =\int\! dx\,
\psi_{\lambda}^{*}(x)\,\psi_{\lambda'}(x)\,\psi_{\lambda''}(x)\,
.
\label{I13}
\end{equation}
The first two integrals of (\ref{I13}) follow from the 
orthonormality relations (\ref{ORTHO}) by taking into account (\ref{COO}):
\begin{equation} \label{I1ab}
I^{\lambda} = \sqrt{M}\, \delta_{\lambda 0} \, ,\quad
I^{\lambda}_{\lambda'} = \delta_{\lambda \lambda'} \, ,
\end{equation}
where $M$ corresponds to $M_T$ or $M_R$, respectively.
Note that we will later make frequently use of the following consequence of (\ref{I13}) and (\ref{I1ab}):
\begin{eqnarray}
\label{CONS}
\hspace*{1mm}\int\! dx\,\psi_{1}(x) =0 \, .
\end{eqnarray}
Integrals with products of more than two eigenfunctions cannot be evaluated 
in general, they have to be determined for each
manifold separately. At present we can only make the following conclusion. Expanding the product
$\psi_{\lambda'}(x)\,\psi_{\lambda''}(x)$ in terms of the complete orthonormal system
\begin{eqnarray}
\psi_{\lambda'}(x)\,\psi_{\lambda''}(x) = 
C_{\lambda' \lambda'' \lambda'''} \psi_{\lambda'''}(x) \, ,
\label{WIGNER}
\end{eqnarray}
the latter integral of (\ref{I13}) is given by
\begin{eqnarray}
I^{\lambda}_{\lambda' \lambda''} =  C_{\lambda' \lambda'' \lambda} \, .
\end{eqnarray}
%
%In a similar way we evaluate the integral (\ref{I4}):
%
%\begin{eqnarray}
%I^{\lambda}_{\lambda' \lambda'' \lambda'''} = 
%C_{\lambda'' \lambda''' \lambda''''} 
%C_{\lambda' \lambda'''' \lambda} \, .
%\end{eqnarray}
%
In addition, we will need also integrals of the type
\begin{equation} 
\label{abkurzneu}
J_{\lambda^{(1)} \lambda^{(2)} \ldots  \lambda^{(n)}}
=\int\! dx\,
\psi_{\lambda^{(1)}} (x)\, \psi_{\lambda^{(2)}} (x)\,\cdots\,
\psi_{\lambda^{(n)}} (x)\,,
\end{equation}
for instance,
\begin{eqnarray}
J_{\lambda \lambda'} &=&\int\! dx\,
\psi_{\lambda} (x)\, \psi_{\lambda'}(x)\,.
\label{I2neu}
\end{eqnarray}
Again we use the orthonormality relations (\ref{ORTHO}), 
the expansion (\ref{WIGNER}), and take into account (\ref{COO}) to
obtain
\begin{eqnarray}
J_{\lambda \lambda'} &=& \sqrt{M} \, C_{\lambda \lambda' 0} \, , 
\end{eqnarray}
where again $M$ corresponds to $M_T$ or $M_R$, respectively.
\subsection{Center Manifold}
Now we approximately determine the center manifold (\ref{H}) 
in lowest order. To this end we read off from (\ref{Q}), (\ref{K}),
and (\ref{S}) that the nonlinear terms
in the equations of evolution for the stable modes $S_{\lambda_T \lambda_R}$
are of quadratic order in the unstable modes $U_{\lambda_T^u \lambda_R^u}$.
Thus, the stable modes can be approximately determined from
\begin{eqnarray}
\dot S_{\lambda_T \lambda_R} = \Lambda_{\lambda_T \lambda_R} \,S_{\lambda_T \lambda_R} + N_{\lambda_T \lambda_R} (U) 
\label{SN}
\end{eqnarray}
with the nonlinearity
\begin{eqnarray}
\label{QN}
N_{\lambda_T \lambda_R} (U) =
\hat P_{\lambda_T \lambda_R} \left( U \hat{C} ( U)
- U \hat{B}(U) - U \hat{B} \left( \hat{C} (U) \right) - 
\hat{B} \left( U \hat{C} (U) \right) \right) \, .
\end{eqnarray}
Using the definitions of the linear operators (\ref{C}), (\ref{B}) and the decomposition 
of the unstable modes (\ref{UEXP}) as well as the projector for the stable modes (\ref{proj2}), 
we see that the second and the third term in (\ref{QN}) vanish due to (\ref{CONS})
\begin{eqnarray}
\hat P_{\lambda_T \lambda_R} \left( U \hat{B}(U) \right) = 
\hat P_{\lambda_T \lambda_R} \left( U \hat{B} \left( \hat{C} (U) \right) \right) = 0 \, ,
\end{eqnarray}
whereas the first term yields
\begin{eqnarray}
\hat P_{\lambda_T \lambda_R} \left(U \hat{C} ( U) \right) =
f_{\lambda_T^u{}'} \, f_{\lambda_R^u{}'} 
\, I^{\lambda_T}_{\lambda_T^u \lambda_T^u{}'} \, 
I^{\lambda_R}_{\lambda_R^u \lambda_R^u{}'}\,
U_{\lambda_T^u \lambda_R^u} \, U_{\lambda_T^u{}' \lambda_R^u{}'} \, , 
\end{eqnarray}
and the fourth term leads to
\begin{eqnarray}
\hat P_{\lambda_T \lambda_R} \left( \hat{B} \left( U \hat{C} (U) \right) \right)& = & 
\frac{1}{2} f_{\lambda_T^u{}'} \, f_{\lambda_R^u{}'} U_{\lambda_T^u \lambda_R^u} \,U_{\lambda_T^u{}' \lambda_R^u{}'} \left[ \frac{1}{\sqrt{M_T}}\, J_{\lambda_T^u \lambda_T^u{}'} \,
I^{\lambda_R}_{\lambda_R^u \lambda_R^u{}'} \,\delta_{\lambda_T 0}\right.\nonumber\\
& & \left.+ \frac{1}{\sqrt{M_R}}\, J_{\lambda_R^u \lambda_R^u{}'} \,I^{\lambda_T}_{\lambda_T^u \lambda_T^u{}'} 
\,\delta_{\lambda_R 0} \right]\,.
\label{FOURTH}
\end{eqnarray}
Therefore, we read off from (\ref{QN})--(\ref{FOURTH}) the decomposition
\begin{eqnarray}
\label{QNN}
N_{\lambda_T \lambda_R}(U) = 
Q^{\lambda_T \lambda_R}_{\lambda_T^u \lambda_R^u,\lambda_T^u{}' \lambda_R^u{}'} \,
U_{\lambda_T^u \lambda_R^u} U_{\lambda_T^u{}' \lambda_R^u{}'} \, ,
\end{eqnarray}
where the expansion coefficients are given by
\begin{eqnarray}
Q^{\lambda_T \lambda_R}_{\lambda_T^u \lambda_R^u,\lambda_T^u{}' \lambda_R^u{}'} &=& 
f_{\lambda_T^u{}'} \, f_{\lambda_R^u{}'} 
\Bigg[ I^{\lambda_T}_{\lambda_T^u \lambda_T^u{}'} \, 
I^{\lambda_R}_{\lambda_R^u \lambda_R^u{}'}
- \frac{1}{2} \, \left( \frac{1}{\sqrt{M_T}}\,J_{\lambda_T^u \lambda_T^u{}'} 
\,I^{\lambda_R}_{\lambda_R^u \lambda_R^u{}'} \,\delta_{\lambda_T 0}\right.\nonumber\\
&&\left.+ \frac{1}{\sqrt{M_R}}\, J_{\lambda_R^u \lambda_R^u{}'} \,I^{\lambda_T}_{\lambda_T^u \lambda_T^u{}'} 
\,\delta_{\lambda_R 0} \right) \Bigg] \,.
\label{QRES}
\end{eqnarray}
Note that Einstein's sum convention is not to be applied. To solve the approximate equations of evolution for the 
stable modes (\ref{SN}) with the quadratic nonlinearity
in the order parameters (\ref{QNN}), we assume that the center 
manifold (\ref{H}) has the same quadratic nonlinearity:
\begin{eqnarray}
\label{HN}
S_{\lambda_T \lambda_R} = 
H^{\lambda_T \lambda_R}_{\lambda_T^u \lambda_R^u,\lambda_T^u{}' \lambda_R^u{}'} \,
U_{\lambda_T^u \lambda_R^u} U_{\lambda_T^u{}' \lambda_R^u{}'} \, .
\end{eqnarray}
Inserting (\ref{HN}) in (\ref{SN}), we only need the linear term in (\ref{U}) to
determine the expansion coefficients of the center manifold:
\begin{eqnarray}
\label{HNN}
H^{\lambda_T \lambda_R}_{\lambda_T^u \lambda_R^u,\lambda_T^u{}' \lambda_R^u{}'}  = \left( 
\Lambda_{\lambda_T^u \lambda_R^u} +\Lambda_{\lambda_T^u{}' \lambda_R^u{}'}
- \Lambda_{\lambda_T \lambda_R}\right)^{-1} 
\,Q^{\lambda_T \lambda_R}_{\lambda_T^u \lambda_R^u,\lambda_T^u{}' \lambda_R^u{}'} \, .
\end{eqnarray}
Here, again, Einstein's sum convention is not to be applied.
Therefore, the Eqs.~(\ref{QRES})--(\ref{HNN}) define the 
lowest order approximation of the center manifold.
\subsection{Order Parameter Equations}
Knowing that the center manifold depends in 
lowest order quadratically on the unstable modes near the instability,
we can determine the order parameter equations up to the cubic 
nonlinearity. Because of (\ref{Q}), (\ref{K}), and (\ref{U}) they read
\begin{eqnarray}
\dot U_{\lambda_T^u \lambda_R^u} = \Lambda_{\lambda_T^u \lambda_R^u} 
\,U_{\lambda_T^u \lambda_R^u} + N_{\lambda_T^u \lambda_R^u} (U,S) \,, 
\label{UN}
\end{eqnarray}
where the nonlinear term decomposes into three contributions:
\begin{eqnarray}
N_{\lambda_T^u \lambda_R^u}(U,S) = Q_{\lambda_T^u \lambda_R^u}(U) +
K_{1,\lambda_T^u \lambda_R^u}(U) + K_{2,\lambda_T^u \lambda_R^u}(U,S) \, . 
\label{UNN}
\end{eqnarray}
The first and the second term represent a quadratic and a 
cubic nonlinearity which is generated by the order parameters themselves
\begin{eqnarray}
Q_{\lambda_T^u \lambda_R^u}(U) & = & \hat P_{\lambda_T^u \lambda_R^u} 
\left( U \hat{C} ( U)
- U \hat{B}(U) - U \hat{B} \left( \hat{C} (U) \right) - \hat{B} 
\left( U \hat{C} (U) \right) \right) \, , 
\label{U1}\\
K_{1,\lambda_T^u \lambda_R^u} (U) & = & - \hat P_{\lambda_T^u \lambda_R^u}
\left( U \hat{B} \left( U \hat{C} (U) \right) \right) \, , 
\label{U2}
\end{eqnarray}
whereas the third one denotes a cubic nonlinearity 
which is affected by the enslaved stables modes according to
\begin{eqnarray}
K_{2,\lambda_T^u \lambda_R^u}(U,S) & = & \hat 
P_{\lambda_T^u \lambda_R^u} \left( U \hat{C} ( S)
- U \hat{B}(S) - U \hat{B} \left( \hat{C} (S) \right) - 
\hat{B} \left( U \hat{C} (S) \right)  \right. \nonumber \\
&& \left. + S \hat{C} ( U)
- S \hat{B}(U) - S \hat{B} \left( \hat{C} (U) \right) - 
\hat{B} \left( S \hat{C} (U) \right) \right) \, .
\label{U3}
\end{eqnarray}
It remains to evaluate the respective contributions by 
using the definitions of the linear operators (\ref{C}), (\ref{B}) and
the decompositions (\ref{UEXP}), (\ref{SEXP}) as well as the 
projector (\ref{proj1}). We start by noting that the
last three terms in (\ref{U1}) vanish due to (\ref{CONS}), i.e.
\begin{eqnarray}
\hat P_{\lambda_T^u \lambda_R^u} \left(U \hat{B}(U) \right) = 
\hat P_{\lambda_T^u \lambda_R^u} \left( U \hat{B} \left( \hat{C} (U) \right) \right) =
\hat P_{\lambda_T^u \lambda_R^u} \left(\hat{B} \left( U \hat{C} (U) \right) \right) = 0 \, ,
\end{eqnarray}
so the first term in (\ref{U1}) leads to the nonvanishing result 
\begin{eqnarray}
\label{QNON}
Q_{\lambda_T^u \lambda_R^u}(U)=
f_{\lambda_T^u{}''} \, f_{\lambda_R^u{}''} \,
I^{\lambda_T^u}_{\lambda_T^u{}' \lambda_T^u{}''} \, 
I^{\lambda_R^u}_{\lambda_R^u{}' \lambda_R^u{}''} \,
U_{\lambda_T^u{}' \lambda_R^u{}'} \,U_{\lambda_T^u{}'' \lambda_R^u{}''} \, .
\end{eqnarray}
Correspondingly, we obtain for (\ref{U2})
\begin{eqnarray}
\hspace{-1cm}K_{1,\lambda_T^u \lambda_R^u}(U) &=& - \frac{1}{2}\,f_{\lambda_T^u{}'''} \, f_{\lambda_R^u{}'''}\, U_{\lambda_T^u{}' \lambda_R^u{}'} \,U_{\lambda_T^u{}'' \lambda_R^u{}''}
\,U_{\lambda_T^u{}''' \lambda_R^u{}'''}\nonumber\\
\hspace{-1cm}&&\hspace{-0.8cm}\times\left( \frac{1}{M_R}\,I^{\lambda_T^u}_{\lambda_T^u{}' \lambda_T^u{}'' \lambda_T^u{}'''} 
\,\delta_{\lambda_R^u \lambda_R^u{}'}
\,J_{\lambda_R^u{}'' \lambda_R^u{}'''} 
+\frac{1}{M_T}\,I^{\lambda_R^u}_{\lambda_R^u{}' \lambda_R^u{}'' \lambda_R^u{}'''} 
\,\delta_{\lambda_T^u \lambda_T^u{}'} 
\,J_{\lambda_T^u{}'' \lambda_T^u{}'''} \right)\,.
\label{K1}
\end{eqnarray}
Furthermore, taking into account (\ref{CONS}), we observe 
that four of the eight terms in (\ref{U3}) vanish:
\begin{eqnarray}
\hspace{-1.2cm}\hat P_{\lambda_T^u \lambda_R^u} \left(S \hat{B}(U) \right)\,, 
\hat P_{\lambda_T^u \lambda_R^u} \left( \hat{B} \left( U \hat{C} (S) \right) \right) \,,
\hat P_{\lambda_T^u \lambda_R^u} \left( S \hat{B} \left( \hat{C} (U) \right) \right) \,,
\hat P_{\lambda_T^u \lambda_R^u} \left( \hat{B} \left( S \hat{C} (U) \right) \right) =  0 \, .
\end{eqnarray}
The nonvanishing terms in (\ref{U3}) read 
\begin{eqnarray}
\hspace*{-0.5cm}\hat P_{\lambda_T^u \lambda_R^u} \left(U \hat{C}(S) \right) &=& 
f_{\lambda_T} \, f_{\lambda_R}\, I^{\lambda_T^u}_{\lambda_T^u{}' \lambda_T} \,
I^{\lambda_R^u}_{\lambda_R^u{}' \lambda_R} \,\, U_{\lambda_T^u{}' \lambda_R^u{}'} S_{\lambda_T \lambda_R} \, , \\
\hspace*{-0.5cm}\hat P_{\lambda_T^u \lambda_R^u} \left(S \hat{C}(U) \right) &=& 
f_{\lambda_T^u{}'} \, f_{\lambda_R^u{}'} \,
I^{\lambda_T^u}_{\lambda_T^u{}'\lambda_T}\, I^{\lambda_R^u}_{\lambda_R^u{}'\lambda_R} \, 
\, U_{\lambda_T^u{}' \lambda_R^u{}'} \, S_{\lambda_T \lambda_R} \, , 
\end{eqnarray}
and
\begin{eqnarray}
\hat P_{\lambda_T^u \lambda_R^u} \left(U \hat{B}(S) \right) &=& - \frac{1}{2}
\left(\frac{1}{\sqrt{M_T}} \,\delta_{\lambda_T 0}\, \delta_{\lambda_T^u \lambda_T^u{}'}
\, I^{\lambda_R^u}_{\lambda_R^u{}'\lambda_R} 
\right.\nonumber\\
&&\left.+\frac{1}{\sqrt{M_R}}\,\delta_{\lambda_R 0}  
\,\delta_{\lambda_R^u \lambda_R^u{}'}\, I^{\lambda_T^u}_{\lambda_T^u{}'\lambda_T}
\right)U_{\lambda_T^u{}' \lambda_R^u{}'} \,S_{\lambda_T \lambda_R} \, , 
\end{eqnarray}
as well as
\begin{eqnarray}
\hat P_{\lambda_T^u \lambda_R^u} \left(U  \hat{B} \left( \hat{C} (S) \right) \right) & = & 
-\frac{1}{2} \,
\left( \,\frac{1}{\sqrt{M_T}}\,\delta_{\lambda_T 0} \, \delta_{\lambda_T^u \lambda_T^u{}'}
 \, f_{\lambda_R} \, 
I^{\lambda_R^u}_{\lambda_R^u{}'\lambda_R} 
\right.\nonumber\\
&&\left.+\frac{1}{\sqrt{M_R}}\,\delta_{\lambda_R 0}\delta_{\lambda_R^u \lambda_R^u{}'} \, f_{\lambda_T} \, 
I^{\lambda_T^u}_{\lambda_T^u{}'\lambda_T}\right)
U_{\lambda_T^u{}' \lambda_R^u{}'} \,S_{\lambda_T \lambda_R} \, ,
\end{eqnarray}
where we used $f_0=1$ in the last equation. Therefore, we obtain for (\ref{U3})
\begin{eqnarray}
&&\hspace{-1cm}K_{2,\lambda_T^u \lambda_R^u} (U,S) =  
U_{\lambda_T^u{}' \lambda_R^u{}'} \,S_{\lambda_T \lambda_R} 
\left\{ \left[ f_{\lambda_T} \, f_{\lambda_R} +f_{\lambda_T^u{}'} \, f_{\lambda_R^u{}'} \, \right] 
I^{\lambda_T^u}_{\lambda_T^u{}' \lambda_T} \,
I^{\lambda_R^u}_{\lambda_R^u{}' \lambda_R} 
\right. \nonumber \\ && \hspace{-1cm}\left.  
- \frac{1}{2} \, \Big[\,\frac{1}{\sqrt{M_T}} \,\delta_{\lambda_T 0} \,\delta_{\lambda_T^u \lambda_T^u{}'} \, 
\left( 1+ f_{\lambda_R} \right) \, I^{\lambda_R^u}_{\lambda_R^u{}' \lambda_R}
+\,\frac{1}{\sqrt{M_R}}\,\delta_{\lambda_R 0}\,\delta_{\lambda_R^u \lambda_R^u{}'}  
\, \left( 1+f_{\lambda_T}\right) \, 
I^{\lambda_T^u}_{\lambda_T^u{}'\lambda_T}
\Big] \right\} \,.
\label{K2}
\end{eqnarray}
Taking into account (\ref{HN}), we read off from (\ref{UN}), (\ref{QNON}), (\ref{K1}), and (\ref{K2}) that
the general form of the order parameter equations is independent
of the geometry of the problem:
\begin{eqnarray}
\dot U_{\lambda_T^u \lambda_R^u} &=&\Lambda_{\lambda_T^u \lambda_R^u} \,
U_{\lambda_T^u \lambda_R^u} + 
A^{\lambda_T^u \lambda_T^u{}' \lambda_T^u{}''}_{\lambda_R^u \lambda_R^u{}' \lambda_R^u{}''} 
\,U_{\lambda_T^u{}' \lambda_R^u{}'} \, U_{\lambda_T^u{}''\lambda_R^u{}''}
\nonumber\\
&&+ B^{\lambda_T^u \lambda_T^u{}' \lambda_T^u{}'' \lambda_T^u{}'''}_{\lambda_R^u{} 
\lambda_R^u{}' \lambda_R^u{}'' \lambda_R^u{}'''} U_{\lambda_T^u{}'\lambda_R^u{}'}
\, U_{\lambda_T^u{}'' \lambda_R^u{}''}\, U_{\lambda_T^u{}''' \lambda_R^u{}'''} \,. 
\label{OPE}
\end{eqnarray}
The corresponding coefficients can be expressed in terms 
of the expansion coefficients $f_{\lambda_T}$, $f_{\lambda_R}$
of the cooperativity functions (\ref{CEX}) and integrals over products of the eigenfunctions  
$\psi_{\lambda_T}(t)$, $\psi_{\lambda_R}(r)$ which have the 
form (\ref{abkurz}) or (\ref{abkurzneu}). They read
\begin{eqnarray}
\label{AA}
A^{\lambda_T^u \lambda_T^u{}' \lambda_T^u{}''}_{\lambda_R^u \lambda_R^u{}' \lambda_R^u{}''}  = f_{\lambda_T^u{}''} \, 
f_{\lambda_R^u{}''} \, I^{\lambda_T^u}_{\lambda_T^u{}' \lambda_T^u{}''} 
\, I^{\lambda_R^u}_{\lambda_R^u{}' \lambda_R^u{}''} \,,
\end{eqnarray}
and
\begin{eqnarray}
&&\hspace{-1.2cm}B^{\lambda_T^u, \lambda_T^u{}' \lambda_T^u{}'' \lambda_T^u{}'''}_{\lambda_R^u{}, \lambda_R^u{}' \lambda_R^u{}'' \lambda_R^u{}'''}
= - \frac{1}{2}\,f_{\lambda_T^u{}'''} \, f_{\lambda_R^u{}'''}
\left( \frac{1}{M_R}\,I^{\lambda_T^u}_{\lambda_T^u{}' \lambda_T^u{}'' \lambda_T^u{}'''}
\,\delta_{\lambda_R^u \lambda_R^u{}'} \,J_{\lambda_R^u{}'' \lambda_R^u{}'''}
+ \frac{1}{M_T}\,I^{\lambda_R^u}_{\lambda_R^u{}' \lambda_R^u{}'' \lambda_R^u{}'''} \delta_{\lambda_T^u \lambda_T^u{}'}
\right.\nonumber\\
&&\hspace{-1.05cm}\times J_{\lambda_T^u{}'' \lambda_T^u{}'''} \bigg)
+\left\{ \left[ f_{\lambda_T} \, f_{\lambda_R} +f_{\lambda_T^u{}'} \, f_{\lambda_R^u{}'} \, \right]
I^{\lambda_T^u}_{\lambda_T^u{}' \lambda_T} \,
I^{\lambda_R^u}_{\lambda_R^u{}' \lambda_R}
- \frac{1}{2} \,
\left[ \,\frac{1}{\sqrt{M_T}}\,\delta_{\lambda_T 0}\, \delta_{\lambda_T^u \lambda_T^u{}'}  \,
\left( 1+ f_{\lambda_R} \right)
I^{\lambda_R^u}_{\lambda_R^u{}' \lambda_R}
\right.\right. \nonumber \\ && \left. \left.
\hspace{-1.05cm}+\,\frac{1}{\sqrt{M_R}}\,\delta_{\lambda_R 0}\,\delta_{\lambda_R^u \lambda_R^u{}'}
\, \left( 1+f_{\lambda_T}\right)
\, I^{\lambda_T^u}_{\lambda_T^u{}' \lambda_T}
\right] \right\} H^{\lambda_T \lambda_R}_{\lambda_T^u{}'' \lambda_R^u{}'',\lambda_T^u{}''' \lambda_R^u{}'''} \,
\,.
\label{BB}
\end{eqnarray}
As is common in synergetics, the coefficients (\ref{BB})
in general consist of two parts, one stemming from the order parameters themselves 
and the other representing the influence of the center manifold $H$.\\

With (\ref{OPE})--(\ref{BB}) we have derived the generic form of the order parameter equations for the connection weights 
between two manifolds of different geometry and dimension.
These equations represent the central new result of our synergetic analysis.
Specifying the geometry means inserting the corresponding eigenfunctions 
of the Laplace-Beltrami operators (\ref{LBO}) into the integrals (\ref{abkurz}), (\ref{abkurzneu}) appearing in (\ref{AA}) and (\ref{BB}). 
Because the synergetic formalism needs not be applied to every geometry anew,
our general procedure means a significant facilitation and tremendous progress as compared to the special approach in Ref.~\cite{Malsburg3}. 

\section{Summary} \label{summary}
In this 
\begin{figure}[t!]
 \centerline{\includegraphics[scale=0.7]{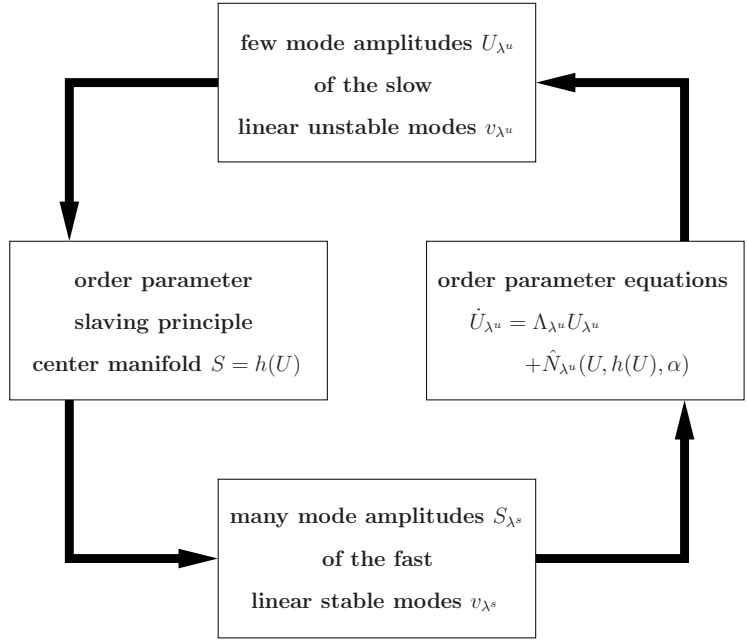}}
 \caption[Circular causality]{\label{zirkaus} \small Circular causality chain of synergetics for the order parameter equations of the generalized H{\"a}ussler equations (\ref{HAUS}). The control parameter $\alpha$ denotes the growth rate of new synapses onto the tectum.}
\end{figure}
paper we have proposed that the self-organized formation of retinotopic projections between manifolds of different geometries and dimensions
is governed by a system of ordinary differential equations (\ref{HAUS}) which generalizes a former ansatz by H{\"a}ussler and
von der Malsburg \cite{Malsburg3}. 
The linear stability analysis determines the instability where an increase of the uniform growth rate $\alpha$ beyond
the critical value (\ref{INS}) converts an initially disordered mapping into a retinotopic projection. Furthermore, it gives rise to a decomposition of the deviation from the stationary uniform solution $v(t,r)$ near the instability in unstable and stable contributions. By inserting this decomposition in the nonlinear H{\"a}ussler equations (\ref{HAUS}), we obtain equations for the mode amplitudes of the unstable and stable modes, respectively. In the vicinity of the instability point the system generates a time-scale hierarchy, i.e.~the stable modes evolve on a faster time-scale than the unstable modes. This leads to the {\it slaving principle} of synergetics: the stable modes are enslaved by the unstable modes. In the literature this enslaving $S=h(U)$ is usually achieved by invoking an adiabatic elimination of the stable modes, which amounts to solving the equation $\dot S=0$. However, the mathematically correct approach for determining the center manifold $h(U)$ is to determine it from the corresponding evolution equations for the stable modes \cite{wwp}. It can be shown that only for real eigenvalues this approach leads to the same result obtained by the approximation $\dot S=0$. Thus, it is 
possible to reduce the original high-dimensional system to a low-dimensional one which only contains the unstable amplitudes. The general form of the resulting order parameter equations (\ref{OPE}) is independent of the geometry of the problem. It contains typically a linear, a quadratic and a cubic term of the order parameters. As a general feature of synergetics, the coefficients (\ref{OPE}), (\ref{BB}) consist of two parts, one stemming from the order parameters themselves and the other representing the influence of the center manifold on the order parameter dynamics.\\

Our results can be interpreted as an example for the validity of the circular causality chain of synergetics, which is illustrated in Figure~\ref{zirkaus}. On the one hand, the order parameters, i.e.~the few amplitudes $U_{\lambda^u}$ of the slowly evolving linear unstable modes $v_{\lambda^u}$, enslave the dynamics of the many stable mode amplitudes $S_{\lambda^s}$ of the fast evolving stable modes $v_{\lambda^s}$ through the center manifold. On the other hand, the center manifold of the stable amplitudes acts back on the order parameter equations. 
\section{Outlook} \label{outlook}

The order parameter equations (\ref{OPE})--(\ref{BB}) represent the central new result of this paper, and in the forthcoming publication \cite{gpw2}
they will serve as the starting point to analyze in detail the self-organization in cell arrays of different geometries. To this end
we assume that the manifolds are characterized by spatial homogeneity and isotropy, i.e.~neither a point
nor a direction is preferred to another, respectively. This additional assumption requires the manifolds to have
a constant curvature and their metric turns out to be the stationary Robertson-Walker metric of general relativity \cite{Weinberg}.
We therefore have to discuss the three different cases where the curvature of the manifolds is positive, vanishes, or is negative.
This corresponds to modelling retina and tectum by the sphere, the plane, or the pseudosphere.\\

A further intriguing problem concerns the question under what circumstances 
non-retinotopic modes become unstable and destroy the retinotopic order. 
One could imagine that some types of pathological development in animals corresponds to this case.\\ 

As already mentioned, lacking any theory for the cooperativity functions, we 
have regarded 
them as time-independent given properties of the manifolds. They are determined
by the lateral connections between the cells of retina and tectum, respectively
\cite{Malsburgskript}. But neither a reason for their time-independence nor a 
detailed discussion of their precise mathematical form is available. To fill 
this gap it will be necessary to elaborate a self-consistent theory of the cooperativity functions.\\

Our generalized H{\"a}ussler equations are fully deterministic. In real systems, 
however, there are always fluctuations. To take into account such unpredictable
small variations a stochastic force has to be added to the deterministic part of the 
equation. Such fluctuations are known to play an important role, especially in 
the vicinity of instability points \cite{Risken,Horsthemke}.\\

Finally, delayed processes could be included in our considerations. Synergetic 
concepts have been successfully applied to time-delayed dynamical systems in Refs.~\cite{wwp,Grigorieva,Simmendinger,sp,sp2}. In neurophysiological systems delays occur due to 
the finite propagation velocity of nerve signals \cite{bpw1,bpw2} as well as the finite duration
of physiological processes such as the change of synaptic connection weights. 
Thus, it would be also worthwhile to expand the investigations to time-delayed H{\"a}ussler equations.

\section*{Acknowledgement}
We thank R. Friedrich, C. von der Malsburg, and A. Wunderlin for stimulating discussions at an initial stage of the work.

\end{document}